\documentclass[a4paper,envcountsame]{llncs}
\usepackage[centering]{geometry}
\usepackage[utf8]{inputenc}
\usepackage{microtype}
\usepackage{amsmath}
\usepackage{amssymb}

\usepackage{amsthm}
\usepackage[colorlinks]{hyperref}
\usepackage{graphicx}
\usepackage{enumitem}

\usepackage{cite}

\usepackage{xcolor}

\usepackage[textsize=small, color=red!30]{todonotes}

\newcommand{\seq}[2][]{#1\langle#2#1\rangle}
\newcommand{\set}[2][]{#1\{#2#1\}}
\newcommand{\pythonlen}[1]{|#1|}

\DeclareMathOperator{\parent}{Parent}
\DeclareMathOperator{\children}{Children}
\DeclareMathOperator{\code}{code}
\DeclareMathOperator{\merge}{merge}
\DeclareMathOperator{\unmerge}{unmerge}

\DeclareMathOperator{\cand}{Candidate-Merges}
\newcommand{\concat}{{}^\frown}
\newcommand{\cycle}[2][]{\seq{\underbrace{\seq{1}, \ldots#1, \seq{1}}_{\text{$#2$~times}}}}

\newcommand{\Components}{\mathcal{C}}
\newcommand{\Graphs}{\mathcal{G}}

\newcommand{\Partitions}{\mathcal{P}}

\newcounter{hypothesisno}

\usepackage{lineno}
\title{Polynomial-delay generation of functional digraphs\\up to isomorphism}

\author{Oscar Defrain \and Antonio E. Porreca \and Ekaterina Timofeeva}
\institute{Aix-Marseille Université, CNRS, LIS, Marseille, France\\
  \email{oscar.defrain@lis-lab.fr}, \email{antonio.porreca@lis-lab.fr}, \email{ekatim239@gmail.com}}

\begin{document}

\maketitle

\thispagestyle{plain}

\begin{abstract}
We describe a procedure for the generation of functional digraphs up to isomorphism; these are digraphs with uniform outdegree~$1$, also called mapping patterns, finite endofunctions, or finite discrete-time dynamical systems. This procedure is based on a reverse search algorithm for the generation of connected functional digraphs, which is then applied as a subroutine for the generation of arbitrary ones. Both algorithms output solutions with $O(n^2)$ delay and require linear space with respect to the number~$n$ of vertices.
% We also provide a proof-of-concept implementation of the algorithms.

\bigskip

\end{abstract}

\section{Introduction and motivation}

A finite, discrete-time dynamical system~$(A,f)$, called in the following just a dynamical system for brevity, is simply a finite set~$A$ of states together with a function~$f \colon A \to A$ describing the evolution of the system in time. A dynamical system can be equivalently described by its transition digraph, which has~$V = A$ as its set of vertices and~$E = \set{(a, f(a)) : a \in A}$ as its set of arcs, that is, each state has an outgoing edge pointing to the next state. Since the dynamical systems we are dealing with are deterministic, their transition digraphs are all and only the digraphs having uniform outdegree~$1$, that is, \emph{functional digraphs}.

The synchronous execution of two dynamical systems~$A$ and~$B$ gives a dynamical system~$A \otimes B$, whose transition digraph is the \emph{direct product}~\cite{Hammack2011a} of the transition digraphs of~$A$ and~$B$. This product, together with a disjoint union operation of sum, gives a semiring structure over dynamical systems up to isomorphism~\cite{Dennunzio2018b} with some interesting algebraic properties, notably the lack of unique factorisation into irreducible digraphs. In order to develop the theory of the semiring of dynamical systems, it is useful to be able to find examples and counterexamples to our conjectures, and this often amounts to being able to efficiently generate all functional digraphs of a given number~$n$ of vertices, up to isomorphism. Remark that the number of non-isomorphic functional digraphs over~$n$ vertices (sequence A001372 on the OEIS~\cite{OEIS2023a}) is exponential, asymptotically $c \times d^n / \sqrt{n}$ for some constants~$c$ and~$d>1$~\cite{Meir1984a}. 
As a consequence, there is no hope to devise an algorithm listing these objects in polynomial time in $n$.
Rather, the kind of efficiency we must aim for is either guaranteeing small exponential time aiming at reducing as much as possible the base of the exponent, referred to as the \emph{input-sensitive} approach \cite{fomin2013exact}, or the ability to generate all solutions in a time which is polynomial in the sizes of the input plus the output, known as the \emph{output-sensitive} approach \cite{johnson1988generating}.
Within this second framework, algorithms running with \emph{polynomial delay} (requiring polynomial time in the input size between consecutive outputs) are considered among the most efficient in algorithmic enumeration.
In this paper, we place ourselves in the output-sensitive approach and refer the reader to \cite{johnson1988generating,strozecki2019survey} for more details on enumeration algorithms and their complexity.

Since the outdegree of each vertex is exactly~$1$ and the number of vertices is finite, the general shape of a functional digraph is a disjoint union of connected components, each consisting of a limit cycle~$\seq{v_1, \ldots, v_k}$, where the vertices~$v_1, \ldots, v_k$ are the roots of~$k$ directed unordered rooted trees (simply referred to as \emph{trees} in the following), with the arcs pointing towards the root.

Enumeration problems for several classes of graphs have been analysed in the literature. For instance, efficient isomorphism-free generation algorithms for rooted, unordered trees are well known, even requiring only amortised constant time per solution~\cite{Beyer1980a}, and there exist polynomial delay algorithms for the isomorphism-free generation of \emph{undirected} graphs~\cite{Goldberg1992a}. More general techniques for the generation of combinatorial objects have been described by McKay~\cite{McKay1998a}. For a practical implementation of generators for several classes of graphs we refer the reader to software such as~\texttt{nauty} and~\texttt{Traces}~\cite{McKay2014a}.

However, the class of functional digraphs does not seem to have been considered yet from the point of view of efficient generation algorithms. Here we first propose a~$O(n^2)$-delay, linear space algorithm for the generation of \emph{connected} $n$-vertex functional digraphs (sequence A002861 on the OEIS~\cite{oeis2023b}), based on an isomorphism code which avoids generating multiple isomorphic digraphs. This assumes the word RAM model with word size~$O(\log n)$~\cite{torben98}.
The algorithm is based on the reverse search technique~\cite{avis1996reverse} that proved to be efficient for the generation of a wide range of objects, including spanning trees of a graph, triangulations in the plane, or cells in a hyperplane arrangement, to cite a few~\cite{avis1996reverse,uno2010efficient,wasa2012constant,conte2016sublinear}. 
In a nutshell, this technique amounts to traverse in a depth-first search manner an implicit solution tree where nodes are solutions, and where edges are defined by some parent-child relation between solutions.
During the traversal, children are obtained by merging trees having adjacent roots along the limit cycle. 
A notable feature of this algorithm is that it can moreover be adapted in order to produce the successor (or predecessor) of any given solution in $O(n^2)$ time as well, and only needs linear space.
This procedure is then used as a subroutine in order to generate \emph{all}, non necessarily connected functional digraphs with the same delay and space.

\section{Isomorphism codes for connected functional digraphs}
\label{sec:codes}

In order to avoid generating multiple isomorphic functional digraphs, we first define a \emph{canonical representation} based on an isomorphism code, which would also allow us to check in polynomial time whether two given functional digraphs are isomorphic when given by another representation (e.g., adjacency lists or matrices).\footnote{Since functional digraphs are planar, they can actually be checked for isomorphism in linear time~\cite{Booth1976a} or in logarithmic space~\cite{Datta2009a}.}

Isomorphism codes for unordered rooted trees (which can be taken as directed with the arcs pointing towards the root, as is needed in our case) are well known in the literature; for instance, \emph{level sequences} (sequences of node depths given by a preorder traversal of the tree, arranged in lexicographic order) can be used for this purpose~\cite{Beyer1980a}. Here we adopt a solution due to 
Jack Edmonds and described by Busacker and Saaty~\cite[pp.~196--199]{Busacker1965a} and Valiente~\cite[p.~118]{Valiente2021a}, which has the useful property that the isomorphism code of a tree directly contains as subsequences the isomorphism codes of its subtrees.

\begin{definition}[code of a tree]
\label{def:tree-code}
Let~$T = (V, E)$ be a tree. Then, the \emph{isomorphism code of~$T$} is the sequence of integers
\begin{align*}
\code{T} = \seq{\pythonlen{V}} \concat \code{T_1} \concat \cdots \concat \code{T_k}
\end{align*}
where~$T_1, \ldots, T_k$ are the immediate subtrees of~$T$, i.e., the subtrees having as roots the predecessors of the root of~$T$, arranged in lexicographically nondecreasing order of code, and~$\concat$ denotes sequence concatenation. In particular, if~$T$ is trivial, i.e., if~$\pythonlen{V} = 1$, then~$\code{T} = \seq{1}$.
\end{definition}

\begin{figure}[t]
    \centering
    \includegraphics[page=1]{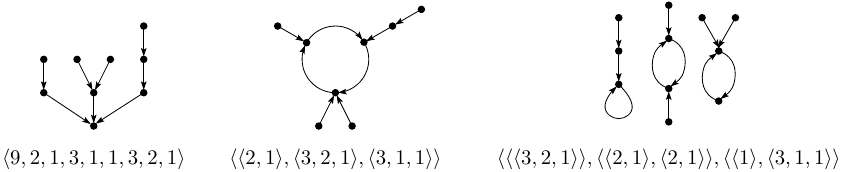}
    \caption{Isomorphism codes for a tree (Definition~\ref{def:tree-code}), a connected (Definition~\ref{def:component-code}) and a disconnected functional digraph (Definition~\ref{def:funcdigraph-code}). Notice the different nesting depths for the angled brackets of the three codes.}
    \label{fig:iso-codes}
\end{figure}

See Fig.~\ref{fig:iso-codes} for an example isomomorphism code for a tree.

For simplicity, in the rest of the paper, we identify a tree with its own code, i.e., we often write~``$T$'' instead of~``$\code{T}$'' where no ambiguity arises. We also denote the lexicographic order on tree isomorphism codes with the symbol~$\le$. As a consequence, a sentence such that~``$T_1 \le T_2$'' is meant to be interpreted as ``the isomorphism code of the tree~$T_1$ is less than or equal to that of~$T_2$ in lexicographic order''. This order has the trivial tree~$\seq{1}$ as its minimum. In the following, we will denote by $\pythonlen{T}$ the length of the code of $T$; note that this value coincides with the number of vertices of $T$ as well as the first integer starting its code.

Since a connected functional digraph consists of a sequence of trees arranged along a cycle, and all rotations of the sequence are equivalent (i.e., they correspond to isomorphic digraphs), we choose a canonical one as its isomorphism code.

\begin{definition}[code of a connected functional digraph]
\label{def:component-code}
The \emph{isomorphism code} of a \emph{connected} functional digraph~$C = (V, E)$ is the lexicographically minimal rotation
\begin{align*}
    \code{C} = \seq{\code T_1, \ldots, \code T_k}
\end{align*}
of the sequence of isomorphism codes of its trees taken in order along the cycle, i.e., such that for all integer~$r$ we have
\begin{align*}
  \seq{\code T_1, \ldots, \code T_k} \le \seq{\code T_{1 + (1 + r) \bmod k}, \ldots, \code T_{1 + (k + r) \bmod k}}.
\end{align*}
\end{definition}

Fig.~\ref{fig:iso-codes} also contains an example of connected functional digraph code.

For brevity, we refer to connected functional digraphs as \emph{components} and, as with trees, we identify a component~$C$ with its own code. A valid code for a component~$C$ is also called a \emph{canonical form} of~$C$; unless otherwise specified, in the rest of the paper we consider all components to be in canonical form. As for trees, we denote the lexicographic order on components (more precisely, on their isomorphism codes) by~$\le$, 
and by $\pythonlen{C}$ the number of trees along the cycle, i.e., the integer $k$ in the definition above.
% and by~$\size{C}$ we denote the number of vertices of component~$C$.
% We call \emph{length} of $C$ and denote by 
% $\pythonlen{C}$ the number of trees along the cycle, i.e., the integer $k$ in the definition above.

Isomorphism codes for arbitrary (i.e., non necessarily connected) functional digraphs will be defined later, in Section~\ref{sec:arbitrary}, since they will be based on the order of generation of components.

Notice that the space required for the isomorphism code of a tree or a connected functional digraph (and later, of an arbitrary functional digraph) of~$n$ vertices is linear on the word RAM model, although the actual number of bits is~$O(n \log n)$, since the codes consist of~$n$ integers ranging from~$1$ to~$n$.

\section{Generation of connected functional digraphs}
\label{sec:connected}

We describe an algorithm based on reverse search~\cite{avis1996reverse} for the enumeration of connected functional digraphs.
This technique is a particular case of a technique called \emph{supergraph method} (or \emph{solution graph traversal}) which amounts to traversing in a depth-first search (DFS) manner an implicit solution graph where nodes are solutions, and edges are defined by some reconfiguration relation between solutions~\cite{johnson1988generating,khachiyan2008enumerating}.
In the framework of \emph{reverse search}, the solution graph is required to be a tree, where the edges are defined by a parent relation.
% This technique \odtext{is a particular case of a technique called \emph{supergraph method} (or \emph{solution graph traversal}) which} amounts to traversing in a depth-first search (DFS) manner an implicit solution tree where nodes are solutions, and edges are defined by some parent-child relation between solutions.
% there is an arc from solution $S$ to solution $S^*$ whenever $S^*=\parent(S)$ for some $\parent$ relation to be defined.
The time and space complexities of the enumeration then essentially boil down to the time and space needed to generate the children of a given node and compute its parent.
When finding a child of a solution, we continue the traversal on this node.
When all children have been found, we backtrack.
When backtracking, we find the next child and continue on this child.
Thus, in general, reverse search only needs memory space that is linear in the height of the solution tree times the space needed to generate children.
As of the delay, it only depends on the time needed to compute the parent and the delay needed to generate children when using folklore tricks such as the alternating output technique~\cite{takeaki2003two}.
We refer the reader to~\cite{avis1996reverse,uno2001fast,uno2010efficient,elbassioni2015polynomial} for more details on this technique. 

% The enumeration of \emph{connected} functional digraphs 
The parent relation we will consider will be based on the following \emph{tree merging} operation, to be applied to the isomorphism codes of two adjacent trees along the cycle.

\begin{definition}[tree merging]
Let~$T_1$ and~$T_2$ be trees with isomorphism codes $\seq{x_1, \ldots, x_k}$ and $\seq{y_1, \ldots, y_\ell}$, respectively. Then
\begin{align*}
    \merge(T_1, T_2) = \seq{x_1 + y_1, x_2, \ldots, x_k, y_1, \ldots, y_\ell}
\end{align*}
that is, the merge is obtained by the concatenation of the codes of the two trees, with the first element updated in order to reflect the new total length.
\end{definition}

The following is an immediate consequence of the definition of the merging operation:

\begin{remark}
If~$T_1$ and~$T_2$ are trees and~$\merge(T_1, T_2)$ is a valid tree isomorphism code, then it represents the tree obtained by connecting~$T_2$ as an immediate subtree of~$T_1$ (i.e., by connecting~$T_2$ to the root of~$T_1$). %\qed
\end{remark}

However, notice that, depending on~$T_1$ and~$T_2$, the result of~$\merge(T_1, T_2)$ is not necessarily a valid tree isomorphism code, as it may not be lexicographically nondecreasing.

\begin{example}
\label{ex:wrong-tree}
Suppose~$T_1 = \seq{4, 3, 2, 1}$ and~$T_2 = \seq{3, 1, 1}$. Then~$\merge(T_1, T_2) = \seq{7, 3, 2, 1, 3, 1, 1}$, which is not a valid isomorphism code, since the subtree code~$\seq{3, 2, 1}$ is lexicographically larger than~$\seq{3, 1, 1}$ which follows it (the actual code for this tree is~$\seq{7, 3, 1, 1, 3, 2, 1}$).
\end{example}

Furthermore, even if the merge of two adjacent trees in a component in canonical form is successful, this does not necessarily produce a valid component isomorphism code either.

\begin{example}
\label{ex:wrong-rotation}
Consider the component $C = \seq{T_1, T_2, T_3} = \seq{\seq{1}, \seq{1}, \seq{1}}$. 
By merging~$T_1$ and~$T_2$ we obtain $\seq{\merge(T_1,T_2), T_3} = \seq{\seq{2,1}, \seq{1}}$
which is not a valid code, since its rotation~$\seq{\seq{1}, \seq{2,1}}$ is strictly inferior in lexicographic order.
\end{example}

Thus, when merging tree codes in a component (in order to produce a new component, as we will describe later) one always needs to check whether the result is indeed a valid code.

We define an inverse of the merging operation on trees that we call \emph{unmerging}.

\begin{definition}[tree unmerging] \label{def:unmerge}Given a tree~$T$, we say that~$\unmerge(T) = (T_1, T_2)$ if and only if~$\merge(T_1, T_2) = T$.
\end{definition}

Notice that there is only one way to unmerge a tree, i.e., detaching the code of its rightmost subtree~$T_2$ (which is already a valid tree isomorphism code) and updating the first element of the remaining tree code in order to reflect the new, shorter length; this is done in linear time and produces a valid tree isomorphism code~$T_1$, since the remaining subtrees are still in nondecreasing order. Thus~$\unmerge$ is indeed a well-defined function. Remark that~$\unmerge(T) = (T_1, T_2)$ implies~$T_1, T_2 < T$, since $T_1,T_2$ are proper subtrees of $T$ and hence have strictly less vertices (their first elements are smaller than the the first element of~$T$).
%
% We can prove that unmerging trees in a component in canonical form in a left-to-right fashion always gives another canonical form, if the two resulting trees are taken in nondecreasing order.\odtodo{C'est le lemme en dessous non ? du coup il faudrait retirer cette phrase}

Notice that a minimal lexicographic rotation always begins with a minimum element of the sequence (with respect to the ordering on elements); more generally, one of the longest subsequences of minimum elements must appear as the prefix of a minimal rotation. Furthermore, a maximal length sequence of one or more minimum elements can never appear as a proper suffix (otherwise, by rotating to the right, we would obtain a smaller rotation); nonetheless, further copies of the same minimum might appear in intermediate positions of the sequence.
These basic observations will be of use in the proof of the next lemma.

% We are now ready to define the aforementioned parent relation between solutions.

\begin{lemma}
\label{thm:unmerge}
Let~$C = \seq{T_1, \ldots, T_k}$ be a component in canonical form with at least one nontrivial tree, and let~$T_h$ be the leftmost such nontrivial tree. Furthermore, let~$(U_1, U_2) = \unmerge(T_h)$ with~$V_1 = \min \set{U_1, U_2}$ and~$V_2 = \max \set{U_1, U_2}$. Then
\begin{align*}
  C^* = \seq{T_1, \ldots, T_{h-1}, V_1, V_2, T_{h+1}, \ldots, T_k}
\end{align*}
is also in canonical form; we call $C^*$ the \emph{parent} of~$C$ and denote it by $\parent(C)$.
\end{lemma}

\begin{proof}
As discussed above, $U_1$ and~$U_2$, and as a consequence $V_1$ and~$V_2$, are always valid tree isomorphism codes.
Thus, we only need to show that~$C^*$ is a minimal rotation. We analyse three sub-cases.

\begin{enumerate}
\item If~$T_1$ is the leftmost nontrivial tree of~$C$ (i.e., if~$h = 1$), then~$C^* = \seq{V_1, V_2, T_{h+1}, \ldots, T_k}$. The tree~$T_1$ is a minimal one in~$C$, and~$V_1 \le V_2 < T_1$, which implies that~$V_1$ and~$V_2$ in the first and second position are both necessary and sufficient for~$C^*$ to be a minimal rotation.

\item If the leftmost nontrivial tree of~$C$ is the last one (i.e., if~$h = k$), then~$C^* = \seq{\seq{1}, \ldots, \seq{1}, V_1, V_2}$. This is always a minimal rotation since~$\seq{1} \le V_1 \le V_2$.

\item Finally, suppose that~$1 < h < k$, i.e., the leftmost nontrivial tree is neither the first nor the last. Then~$C^*$ begins with a sequence of trivial trees of length~$h-1$, $h$, or~$h+1$, depending on whether~$V_1$ and~$V_2$ are trivial, followed by a nontrivial tree (resp., $V_1$, $V_2$, or~$T_{h+1}$).
We distinguish cases accordingly:

\begin{itemize}
\item If~$V_1$ or both~$V_1$ and~$V_2$ are trivial (notice that~$V_2$ trivial implies~$V_1$ trivial, since~$V_1 \le V_2$) the maximal prefix of trivial trees of~$C^*$ has length strictly larger than~$h-1$. If~$C^*$ is not a minimal rotation, the actual minimal one would include a sequence of at least~$h$ trivial trees as a prefix, to be found as a subsequence of~$\seq{T_{h+1}, \ldots, T_{k-1}}$. But this sequence would also exist in~$C$ and would lead to a smaller minimal rotation, since~$C$ begins with only~$h-1$ trivial trees, contradicting the hypothesis that~$C$ is a canonical form.

\item If both~$V_1$ and~$V_2$ are nontrivial, then~$C^*$ has a prefix of trivial trees of length~$h-1$, followed by~$V_1$. Suppose, by contradiction, that~$C^*$ is not a minimal rotation. Then, a sequence of length~$h$ of the form~$P = \seq{\seq{1}, \ldots, \seq{1}, T}$ exists, as a subsequence of~$\seq{T_{h+1}, \ldots, T_k}$, with the property that~$P \le \seq{T_1, \ldots, T_{h-1}, V_1}$, i.e., $P$~is the prefix of the actual minimal rotation ($P$~and the prefix of~$C^*$ could be elementwise identical, the difference only appearing later). This implies that~$T \le V_1$. But~$V_1 \le T_h$ and thus~$P \le \seq{T_1, \ldots, T_h}$; since~$P$ is also a subsequence of~$C$, this implies that~$C$ is not a minimal rotation, once again a contradiction. \qedhere
\end{itemize}
\end{enumerate}
\end{proof}

A key property given by Lemma~\ref{thm:unmerge} is the following.

\begin{proposition}\label{prop:tree}
If $C,C^*$ are two components such that $C^*=\parent(C)$ then $\pythonlen{C^*}=\pythonlen{C}+1$.
\end{proposition}

\begin{proof}
This is a consequence of the fact that the parent of a component is obtained by unmerging its leftmost nontrivial tree, and hence increases its number of trees by exactly one.
\end{proof}

Note that the $n$-vertex simple cycle is the unique $n$-vertex component maximising the length of its code, which is $n$.
As a consequence, together with Proposition~\ref{prop:tree}, it is easily proved that the $\parent$ relation defines a rooted tree of height at most $n-1$, having as nodes the $n$-vertex components, with the $n$-vertex cycle as the root, and with an edge between two components $C$ and $C^*$ whenever $C^*=\parent(C)$.
It now suffices to show how to generate all $C$ such that $C^*=\parent(C)$ for a given component $C^*$, in order to obtain an efficient generation algorithm by reverse search:
all solutions will be obtained by a traversal of the solution tree initiated at the root being the $n$-vertex simple cycle, and duplicates are inherently avoided by the acyclicity of the solution tree; see e.g.,~\cite{elbassioni2015polynomial} for a formalisation of the technique.
In the following, we shall note $\children(C^*)$ the family of such components~$C$.
In order to list this family, we define the set of \emph{candidate merges} of a given component, that we will show to contain every child.

\begin{definition}
\label{def:merges}
Let~$C^* = \seq{T_1, \ldots, T_k}$ be a component with $k\geq 2$ and let
\begin{align*}
    L_{i}(C^*) &= \seq{T_1, \ldots, T_{i-1}, \merge{(T_i, T_{i+1})}, T_{i+2}, \ldots, T_k},\\
    R_{i}(C^*) &= \seq{T_1, \ldots, T_{i-1}, \merge{(T_{i+1}, T_i)}, T_{i+2}, \ldots, T_k},
\end{align*}
for~$1 \le i < k$. 
Then, we denote the set of candidate merges of~$C^*$ by
\begin{align*}
  \cand(C^*) = \bigcup_{1\leq i<k} \set[\big]{L_{i}(C^*), R_{i}(C^*)}.
\end{align*}
\end{definition}

% Notice that, since~$\cunmerge$ is a function, for distinct components~$C_1$ and~$C_2$ we always have~$\merges C_1 \cap \merges C_2 = \varnothing$.
The candidate merges of~$C^*$ thus consist of all components obtained by merging each pair of adjacent trees in both directions.
We note that the candidate merges of two distinct components could intersect, a point that will not be important in the following as we will only explore actual children among these candidates; for instance, both~$\seq{\seq{1}, \seq{1}, \seq{1}, \seq{3,2,1}}$ and~$\seq{\seq{1}, \seq{2,1}, \seq{1}, \seq{2,1}}$ would share an element among their candidate merges:
\begin{align*}
    \seq{\seq{1}, \seq{2,1}, \seq{3,2,1}} &= \seq{\seq{1}, \merge(\seq{1}, \seq{1}), \seq{3,2,1}} \\
    &= \seq{\seq{1}, \seq{2,1}, \merge (\seq{1}, \seq{2,1})}.
\end{align*}

% \begin{lemma}
% Let~$C = \seq{T_1, \ldots, T_k}$ be a component. For all~$i = 1, \ldots, k$ such that~$T_i$ is nontrivial, let
% \begin{linenomath*}
% \begin{align*}
%   C_i = \seq{T_1 \ldots, T_{i-1}} \concat \unmerge{T_i} \concat \seq{T_{i+1}, \ldots, T_k}
    %   \end{align*}
% \end{linenomath*}
% Let~$T_h$ be the leftmost nontrivial tree of~$C$. Then
% \begin{linenomath*}
    %     \begin{align*}
%   C_h = \min \set{C_i : C_i \text{ is in canonical form}}.
    %   \end{align*}
% \end{linenomath*}
% \end{lemma}

% \begin{proof}
% We already know that~$C_h$ is in canonical form by Lemma~\ref{thm:unmerge}. Since~$T_h$ is the leftmost nontrivial tree of~$C$ and since the first tree of~$\unmerge{T_h}$ is trivial, we know that~$C_h$ begins with a sequence of at least~$h$ trivial trees. For all~$i > h$, on the other hand, $C_i$ begins with only~$h-1$ trivial trees, followed by the nontrivial~$T_h$. But then~$C_h$ is lexicographically strictly less than~$C_i$.
% \end{proof}

\begin{lemma}
\label{thm:min-rotation-linear}
Given~$C = \seq{T_1, \ldots, T_k} \in \cand(C^*)$ for some component~$C^*$, it is possible to check in~$O(n)$ time, where~$n = \pythonlen{T_1} + \cdots + \pythonlen{T_k}$, if~$C$ is a valid isomorphism code for a component.
\end{lemma}

\begin{proof}
First of all, it is necessary to check if each~$T_i$ is a valid isomorphism code for a tree. Since either~$T_i = \seq{1}$, or it is obtained by merging two valid tree codes, it suffices to check that the codes of the subtrees of~$T_i$ are in lexicographic order (see Example~\ref{ex:wrong-tree}). The root of each subtree is the length of the corresponding subsequence, hence the subtrees can be extracted in linear time, and the resulting sequence of tree codes can also be checked for lexicographic order in linear time.

If the previous test is satisfied for all~$T_i$, then it suffices to check if~$C$ is a lexicographically minimal rotation (see Example~\ref{ex:wrong-rotation}). On ``flat'' sequences of integers, linear time algorithms such as Booth's~\cite{Booth1980a,Booth2019a} are well known. We can solve our problem (which involves sequences of sequences of integers) by reduction to this simpler case.

Let~$P = \seq{0} \concat T_1 \concat \seq{0} \concat T_2 \concat \cdots \concat \seq{0} \concat T_k$ be the concatenation of the codes of the given trees, each prefixed with an extra~$\seq{0}$; remark that $0$ is strictly less than any integer appearing in the code of a tree. Let~$P'$ be the minimal rotation of~$P$; then~$P'$ must begin with 0, since it is the minimum element of the sequence, and thus~$P' = \seq{0} \concat U_1 \concat \seq{0} \concat U_2 \concat \cdots \concat \seq{0} \concat U_k$, where each~$U_i$ is one of the original trees~$T_j$ and~$\seq{U_1, \ldots, U_k}$ is a rotation of~$\seq{T_1, \ldots, T_k}$. 

We claim that~$\seq{U_1, \ldots, U_k}$ is, more specifically, the \emph{minimal} rotation of $\seq{T_1, \ldots, T_k}$. Otherwise, by contradiction, there would exist another rotation $\seq{V_1, \ldots, V_k} < \seq{U_1, \ldots, U_k}$. Let~$V_i$ and $U_i$ be the leftmost trees such that~$V_i \ne U_i$. Then
\begin{align*}
    \underbrace{\seq{0} \concat V_1 \concat \cdots \concat \seq{0}}_{V} \concat V_i \concat \cdots \concat \seq{0} \concat V_k <
    \underbrace{\seq{0} \concat U_1 \concat \cdots \concat \seq{0}}_{U} \concat U_i \concat \cdots \concat \seq{0} \concat U_k = P'
\end{align*}
since the two prefixes~$U$ and~$V$ are identical and~$V_i < U_i$. But this contradicts the assumption that~$P'$ is the minimal rotation of $P$.

Then $P$ is its own minimal rotation if and only if $\seq{T_1, \ldots, T_k}$ is a minimal rotation. Since~$P$ is a sequence of integers, the former property can be checked in linear time as mentioned above.
\end{proof}

We are now ready to describe how to efficiently compute the children of a component.

\begin{lemma}\label{lem:children-obtained}
Let $C^*$ be a component.
Then every $C\in \children(C^*)$ is a candidate merge of $C^*$.
Furthermore, the set $\children(C^*)$ can be enumerated with delay $O(n^2)$ and linear space.
\end{lemma}

\begin{proof}
Let us first assume that $\children(C^*)$ is nonempty and let $C\in \children(C^*)$. 
Suppose that $C=\seq{T_1,\dots,T_k}$ for some integer $k\geq 1$.
By definition, $C^* = \seq{T_1, \ldots, T_{h-1}, V_1, V_2, T_{h+1}, \ldots, T_k}$ for $T_h$ the leftmost nontrivial tree of $C$, and where~$(U_1, U_2) = \unmerge(T_h)$, $V_1 = \min \set{U_1, U_2}$, and~$V_2 = \max \set{U_1, U_2}$.
By definition of $\unmerge$, we have that at least one of
\[ 
    C= \seq{T_1, \ldots, T_{h-1}, \merge(U_1, U_2), T_{h+1}, \ldots, T_k}
\]
or
\[ 
    C= \seq{T_1, \ldots, T_{h-1}, \merge(U_2, U_1), T_{h+1}, \ldots, T_k}
\]
holds depending on whether $U_1$ or $U_2$ is minimal.
We conclude that at least one of $L_i(C^*)$ or $R_i(C^*)$ equals $C$ for $i=h$, hence that $C$ belongs to $\cand(C^*)$.
We note that $L_i(C^*)=R_i(C^*)$ whenever $U_1=U_2$;
even though this has no incidence in the proof as $\cand(C^*)$ is defined as a set, one should take care of this situation by considering only one of $L_i(C^*)$ or $R_i(C^*)$ in the implementation.

Regarding the delay, note that the number of merge operations to perform in order to generate $\cand(C^*)$ is $2(k-1)$, and that each $\merge$ takes $O(n)$ time.
When performing a $\merge$, we additionally check whether the obtained code is valid, which can be done in~$O(n)$ time by Lemma~\ref{thm:min-rotation-linear}, and whether it is a child of $C^*$, by performing one $\unmerge$ operation once again in~$O(n)$ time.
Since $k\leq n$, we obtain a $O(n^2)$-delay algorithm generating all children of $C^*$ as desired. The space is linear since we only store a constant number of components.

Now if $\children(C^*)$ is empty, then the first part of the statement trivially holds, and the enumeration amounts to decide whether $\children(C^*)=\varnothing$ which is done by seeking for children as described above, in $O(n^2)$ time and linear space as well.
\end{proof}

Let us mention for completeness that situations can occur where for a given component $C^*$ and integer $i$, each of the components $L_i(C^*)$ and $R_i(C^*)$ define a distinct child; for example, this happens for $C^*=\seq{\seq{2,1}, \seq{3,2,1}}$, $n=5$ and $i=1$.
On the other hand, root-to-leaf paths of length less than $n-1$ can occur in the solution tree, with components not having any children despite being of length more than 1; an example is given by $C^*=\seq{\seq{2,1}, \seq{2,1}, \seq{2,1}}$ for $n=6$.
In situations of this kind, every valid candidate merge is actually the child of another component.

We note that the general framework of reverse search~\cite{avis1996reverse}, equipped with the alternating output technique~\cite{takeaki2003two}, yields a natural polynomial-time algorithm to produce the solution that comes after any given solution $C$ in the enumeration, provided that we are able to decide in polynomial time whether $C$ lies at even or odd depth in the solution tree, and to resume the children enumeration of an arbitrary node at child $i$ for any integer $i$. 
This comes from the fact that the next solution is at distance at most 3 from $C$ in the solution tree, hence that resuming the DFS on $C$, we obtain the next solution after at most 3 recursive calls and/or backtracks, in a time which is bounded by 3 times the delay needed to produce the next child. 
In our case, as the worst case delay to generate a child is the same as of generating all children, we do not even need to argue that we can resume the enumeration from the $i$-th child as long as we are interested in the asymptotic delay, though this can be done to speed up the implementation.
As of deciding whether the current solution $C$ lies at even or odd depth in the solution tree, it can be done by checking the parity of $n-\pythonlen{C}$, that is, the number of merges performed starting from the root in order to obtain~$C$.
By the same arguments, it is easily seen that we can compute the solution that comes right before any given solution $C$ within the same time and space.
In other words, our implementation of reverse search can be made so that it only uses the memory of the current node during the DFS of the solution tree, within the same delay.
We conclude to the following.

\begin{theorem} 
\label{thm:generates-all-connected}
There is a~$O(n^2)$-delay and linear space algorithm generating all connected \mbox{$n$-vertex} functional digraphs. 
Moreover, given any such functional digraph, we can generate its successor (resp., predecessor) in the enumeration in $O(n^2)$ time and using linear space. \qed
\end{theorem}

\begin{figure}[t]
    \centering
    \includegraphics[page=2]{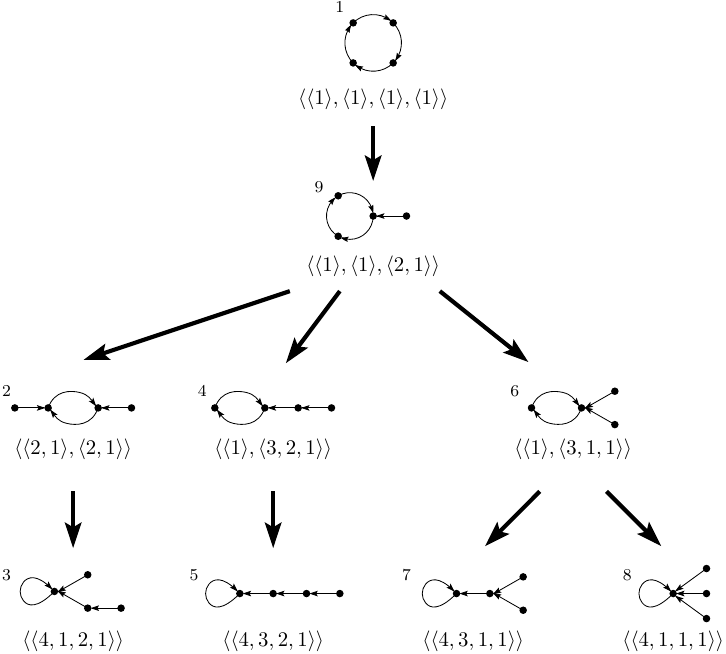}
    \caption{Reverse search tree for the generation of components of~$4$ vertices, represented both in graphical form (top) and as isomorphism codes (bottom); the order of generation by the algorithm of Theorem~\ref{thm:generates-all-connected} is displayed on the top left. Note that the actual ordering of children of each node depends on the (arbitrary) order we chose for the generation of the candidate merges; in the picture we first compute $L_i(C^*)$ for $i$ from $1$ to $k-1$, then the same for the $R_i(C^*)$: this is the ordering that has been adopted in our implementation and that will be considered in the remaining of the paper.}
    \label{fig:tree}
\end{figure}

As an example, the generation of all components of~$4$ vertices is depicted in Fig.~\ref{fig:tree}.

\section{Generation of arbitrary functional digraphs}
\label{sec:arbitrary}

We can now exploit the algorithms of Theorem~\ref{thm:generates-all-connected}, and more specifically our ability to generate the successor of a given component, as a subroutine for the efficient generation of arbitrary (non necessarily connected) functional digraphs. In order to avoid generating multiple isomorphic digraphs, we first define an appropriate isomorphism code.

% We can now exploit Algorithm~\ref{alg:connected-no-repetitions} as a subroutine in order to devise an efficient algorithm for the generation of arbitrary (non necessarily connected) functional digraphs. More precisely, like Algorithm~\ref{alg:connected-no-repetitions}, we will be able to compute the successor of a given functional digraph under some ordering. As a subroutine, we exploit an algorithm for generating partitions of an integer~$n$~\cite{Kelleher2009a}.

% An arbitrary functional digraph can be represented by a sequence of its connected components. Since there is no intrinsic order among components, any permutation of the sequence would represent the same functional digraph; as a consequence, once again we choose a canonical permutation.

% An arbitrary functional digraph is the (possibly empty) disjoint union of components, up to isomorphism. Equivalently, a functional digraph can be seen as a multiset of components up to isomorphism. Another characterisation, possibly more useful for enumeration purposes, is the following: for each~$n \ge 1$ let~$\Components_n$ be the set of components of~$n$ vertices, and for each~$n \ge 0$ let~$\Graphs_n$ be the set of functional digraphs of~$n$ vertices, both up to isomorphism. Furthermore, for each set~$X$ and~$n \ge 0$, let~$\Multisets_n(X)$ be the set of multisets of~$n$ elements of~$X$. Then
% \begin{align*}
%     \Graphs_n = \bigcup_{p \in \Partitions_n} \prod_{1 \le i \le n} \Multisets_{p_i}(\Components_i)
% \end{align*}

\begin{definition}[code of a functional digraph]
\label{def:funcdigraph-code}
Let~$G = (V, E)$ be an arbitrary functional digraph having~$m$ connected components. Then, the \emph{isomorphism code of~$G$} is the sequence
\begin{align*}
  \code{G} = \seq{\code{C_1}, \ldots, \code{C_m}}.
\end{align*}
where, for each pair of consecutive components~$C_i$ and~$C_{i+1}$, either~$C_i$ has less vertices than $C_{i+1}$, or~both $C_i$ and $C_{i+1}$ have the same number of vertices and their codes are in increasing order of generation by the algorithm of Theorem~\ref{thm:generates-all-connected}.
\end{definition}

An example code for a disconnected functional digraph is depicted in Fig.~\ref{fig:iso-codes}.

As usual, we identify a functional digraph~$G$ with its own code and refer to it as a canonical form for~$G$. %, and we denote its number of vertices by~$\size{G}$.

Notice how the isomorphism code for a functional digraph resembles a PQ-tree, a data structure representing permutations of a given set of elements which, incidentally, is used to efficiently check isomorphic graphs of certain classes~\cite{Booth1976a}. However, for our application we need to represent the equivalence of all permutations of the components~$C_1, \ldots, C_m$, as well as the equivalence of all rotations of the trees~$T_1, \ldots, T_k$ of a component~$C_i$, and this latter condition is not represented directly in a PQ-tree.

\begin{definition}
We denote by~$\Components_n$ the set of components of~$n$ vertices, and by~$\Components_n^m$ the set of functional digraphs having~$m$ components of~$n$ vertices each.
\end{definition}

In terms of isomorphism codes, the elements of~$\Components_n^m$ have the form~$\seq{C_1, \ldots, C_m}$ with all~$C_i \in \Components_n$ and~$C_1 \le \cdots \le C_m$ in generation order. Remark that~$\Components_n^0 = \set{\seq{}}$ as a base case, i.e., it only contains the empty functional digraph, and that~$\Components_n^m$ is not simply the Cartesian product of~$m$ copies of~$\Components_n$, due to the ordering requirement, but rather the set of \emph{multisets} of~$m$ elements of~$\Components_n$. This set can be generated efficiently as follows.

\begin{lemma}
\label{thm:generating-multisets}
$\Components_n^m$ can be generated with delay~$O(m n^2)$ and in space~$O(mn)$.
\end{lemma}

\begin{proof}
We generate~$\Components_n^m$ starting from the functional digraph
\begin{align*}
  G_0 = \seq{\underbrace{\cycle{n}, \ldots, \cycle{n}}_{m \text{ times}}}
\end{align*}
that is, the functional digraph consisting of~$m$ cycles of length~$n$, which are the first elements (the roots of the solution trees) in the generation order of components.

Now let~$G = \seq{C_1, \ldots, C_m}$ be an arbitrary element of~$\Components_n^m$. The successor~$G'$ of~$G$ is computed by taking the rightmost element~$C_i$ that possesses a successor component~$C_i'$, if any, and letting
\begin{align*}
  G' = \seq{C_1, \ldots, C_{i-1}, \underbrace{C_i', \ldots, C_i'}_{m-i+1 \text{ times}}}
\end{align*}
that is, by replacing~$C_i$ and every other component on its right by~$C_i'$.

The resulting functional digraph code~$G'$ is still in nondecreasing order of generation of components, thus a valid isomorphism code, and~$G' > G$ in the induced lexicographical order, which avoids generating the same code multiple times. Furthermore, any valid functional digraph code can be constructed right-to-left from~$G_0$ by applying the successor operation enough times to the cycle of length~$n$. This is thus an exhaustive generation without duplicates. Intuitively, this amounts to generating all nondecreasing words of length~$m$ on the alphabet~$\Components_n$, where nondecreasing is to be interpreted with respect to the order of generation of~$\Components_n$.

Computing~$G'$ requires applying the successor operation to at most~$m$ components, and replicating at most~$m$ times the successor component~$C_i'$ thus obtained, which can be carried out in~$O(mn^2)$ time. The space bound is due to the fact that we only store one functional digraph.
\end{proof}

Recall that an \emph{integer partition} of a natural number~$n$ is an (unordered) multiset of positive integers having sum~$n$. Partitions can be alternatively represented as a vector of~$n$ natural numbers (including zero) representing the multiplicities of each term of the sum.

\begin{definition}
\label{def:partition}
A partition of the natural number~$n$ is an~$n$-tuple~$p = (p_1, \ldots, p_n)$ of natural numbers such that~$\sum_{i=1}^n p_i i = n$. We denote by~$\Partitions_n$ the set of partitions of~$n$.
\end{definition}

\begin{definition}
We denote by~$\Graphs_n$ the whole set of functional digraphs over~$n$ vertices. Furthermore, for each partition~$p = (p_1, \ldots, p_n) \in \Partitions_n$, we denote by~$\Graphs_p$ the set of functional digraphs having exactly~$p_i$ components of~$i$ vertices for all~$1 \le p \le n$.
\end{definition}

By definition we have~$\Graphs_n = \bigcup_{p \in \Partitions_n} \Graphs_p$, and $\Graphs_p$ and~$\Graphs_q$ are disjoint whenever~$p$ and~$q$ are are distinct partitions of~$n$; in other words, $\set{\Graphs_p : p \in \Partitions_n}$ is a (set) partition of~$\Graphs_n$. Since integer partitions can be generated efficiently,\footnote{With a suitable representation of partitions, an amortised constant delay is even possible~\cite{kelleher2009a}.} we only need to show that each~$\Graphs_p$ can be generated efficiently. By ``grouping together'' all components of the same size, we can analyse an arbitrary functional digraph~$G \in \Graphs_p$ as the disjoint union of (possibly empty) digraphs~$G_i \in \Components_i^{p_i}$ for~$1 \le i \le n$. Since the components of~$G_i$ are smaller than those of~$G_{i+1}$ for all~$1 \le i < n$, we have~$G = G_1 \concat \cdots \concat G_n$ in terms of isomorphism codes.

\begin{theorem}
The set of functional digraphs over~$n$ vertices up to isomorphism can be generated with delay~$O(n^2)$ and using linear space.
\end{theorem}

\begin{proof}
Since integer partitions can be generated with linear delay~\cite{kelleher2009a}, it suffices to show that each set~$\Graphs_p$ can be generated with delay~$O(n^2)$ in order to prove the theorem.

The set~$\Graphs_p$~can be generated by enumerating the $n$-tuples~$\seq{G_1, \ldots, G_n}$ of the Cartesian product~$\prod_{i=1}^n \Components_i^{p_i}$ and then outputting~$G_1 \concat \cdots \concat G_n$. Generating the next element of a Cartesian product may require, in the worst case, trying to compute the successor of each coordinate~$\Components_i^{p_i}$, and resetting that coordinate to the first element of~$\Components_i^{p_i}$ (that is, the functional digraph consisting in~$p_i$ cycles of length~$i$) if there is no successor. By Lemma~\ref{thm:generating-multisets}, each~$\Components_i^{p_i}$ can be generated with delay~$O(p_i i^2)$, which means that the generation delay for~$\Graphs_p$ is asymptotically proportional to~$\sum_{i=1}^n p_i i^2 \le n \sum_{i=1}^n p_i i$. By recalling that~$\sum_{i=1}^n p_i i = n$ (Definition~\ref{def:partition}) we obtain the required quadratic upper bound on the delay. Since we never store more than one functional digraph at a time, and since the successor of each component can be computed in linear space (Theorem~\ref{thm:generates-all-connected}), the overall space requirement for this algorithm is also linear.
\end{proof}

\begin{figure}[t]
    \centering
    \includegraphics[page=3]{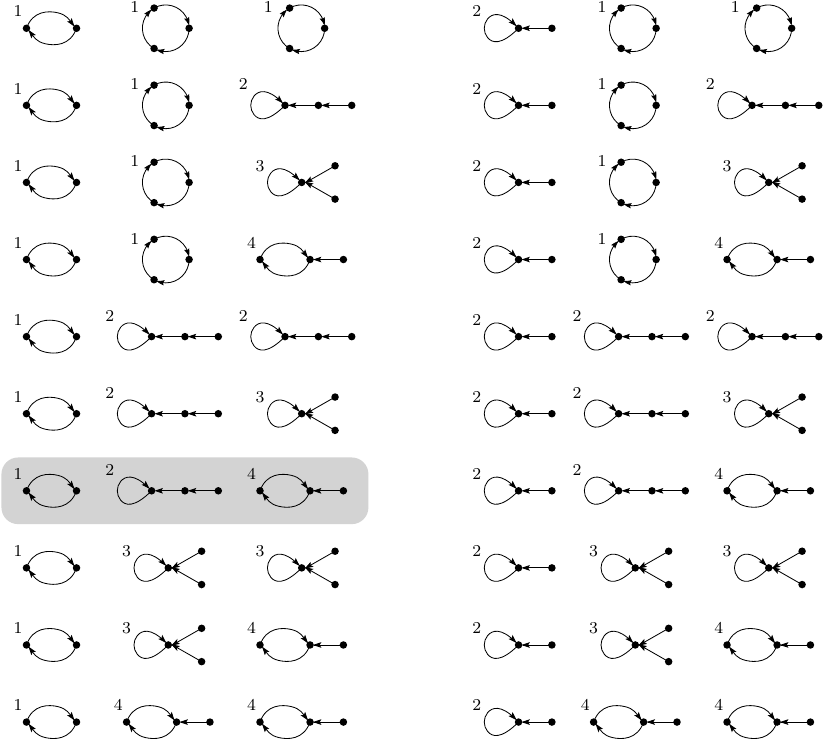}
    \caption{Generation (top to bottom, left to right) of the set of functional digraphs~$\Graphs_p$ over~$8$ vertices with partition~$p = (0, 1, 2, 0, 0, 0, 0, 0)$, corresponding to components of~$2$, $3$, $3$ vertices respectively. Each component is labelled by its generation order according to the algorithm of Theorem~\ref{thm:generates-all-connected}. As an example, the functional digraph in grey (the 7th generated one in~$\Graphs_p$) has isomorphism code~$\seq{\seq{\seq{1},\seq{1}}, \seq{\seq{3,2,1}}, \seq{\seq{1},\seq{2,1}}}$.}
    \label{fig:gen-funcdigraphs}
\end{figure}

Fig.~\ref{fig:gen-funcdigraphs} shows an example generation of disconnected functional digraphs with a fixed partition of vertices into components, and highlights the similarity with the increment operation on a mixed radix integer (with the extra constraint that components of the same size must be in nondecreasing order of generation).

\section{Conclusions}
\label{sec:conclusions}

We have described the first polynomial-delay generation algorithm for the class of functional digraphs, both connected and arbitrary, which proves that these classes of graphs can be generated with an~$O(n^2)$ delay and linear space.\footnote{A proof-of-concept implementation of the algorithms described in this paper, the~\texttt{funkdigen} command-line tool, is also available~\cite{Porreca2023b}.}

It is, of course, an open problem to establish if functional digraphs can be generated with a smaller delay. That would require us to somehow avoid testing~$O(n)$ possible merges in order to construct the next candidate digraph, or to avoid spending linear time in order to check for valid isomorphism codes. Otherwise, is an amortised constant time delay possible?

Another interesting line of research is to find variations of the tree-merging approach suitable for the efficient generation either of restricted classes of functional digraphs (for instance, with cycles of given lengths or trees of given heights, which is sometimes useful in applications related to the decomposition of dynamical systems~\cite{Dennunzio2018b}), or of more general classes of graphs, without the uniform outdegree~$1$ constraint, possibly by means of a ``functional digraph decomposition''.

\subsubsection*{Acknowledgements}

Ekaterina Timofeeva was funded by the undergraduate internship program ``Incubateurs de jeunes scientifiques'' of the Faculty of Sciences at Aix-Marseille Université and by the Agence National de la Recherche project ANR-18-CE40-0002 FANs.

We would like to thank Kellogg S. Booth, Jerome Kelleher, Brendan D. McKay, and K\'evin Perrot for some useful discussions and for providing some bibliographic references. We are also grateful to an anonymous reviewer for suggesting the explicit use of the supergraph method, which allowed us to simplify the proofs and to identify a bottleneck in a previous version of our component generation algorithm, thus improving its efficiency.

\bibliographystyle{splncs04}
\bibliography{Bibliography}

\end{document}